\documentclass{nature}

\usepackage{amssymb}
\usepackage{graphicx}
\usepackage{bm}
\usepackage{amsmath}
\usepackage{gensymb}
\usepackage[utf8]{inputenc}
\usepackage{color}
\usepackage{amssymb}
\usepackage{ulem}
\usepackage{lineno}

\usepackage{xpatch} 

\title{A measure of the size of the magnetospheric accretion region in TW Hydrae.}

\author{GRAVITY Collaboration: R. Garcia Lopez$^{1,2,3}$, A. Natta$^{2}$, A. Caratti o Garatti$^{1,2,3}$,  T.P. Ray$^{2}$, R. Fedriani$^{2,16}$, M. Koutoulaki$^{2,7}$, L. Klarmann$^{3}$, K. Perraut$^{15}$, J. Sanchez-Bermudez$^{3,18}$, M. Benisty$^{15,13}$, C. Dougados$^{15}$, L. Labadie$^{4}$, W. Brandner$^{3}$, P.J.V. Garcia$^{5,6,10}$, Th. Henning$^{3}$, P. Caselli$^{8}$, G.~Duvert$^{15}$, T. de Zeeuw$^{8,14}$, R. Grellmann$^{4}$, R.~Abuter $^{7}$, A.~Amorim $^{6,17}$, M.~Baub\"{o}ck${^8}$, J.P.~Berger $^{7,15}$, H.~Bonnet $^{7}$, A.~Buron $^{8}$,  Y.~Cl\'enet $^{9}$, V.~Coud\'e~du~Foresto$^{9}$, W.~de~Wit$^{10}$, A.~Eckart$^{4,11}$, F.~Eisenhauer$^{8}$, M.~Filho$^{5,6,10}$, F.~Gao${^8}$, C.E.~Garcia Dabo$^{7}$, E.~Gendron$^{9}$, R.~Genzel $^{8,12}$,  S.~Gillessen$^{8}$, M. Habibi${^8}$, X.~Haubois$^{10}$, F.~Haussmann$^{8}$, S.~Hippler$^{3}$, Z.~Hubert$^{15}$, M.~Horrobin$^{4}$, A. Jimenez Rosales${^8}$, L.~Jocou$^{15}$, P.~Kervella$^{9}$, J.~Kolb$^{10}$, S.~Lacour$^{9}$, J.-B.~Le~Bouquin$^{15}$, P.~L\'ena$^{9}$, T.~Ott$^{8}$, T.~Paumard$^{9}$, G.~Perrin$^{9}$, O.~Pfuhl$^{7}$, A.~Ramirez$^{7}$, C.~Rau$^{8}$, G.~Rousset$^{9}$, S.~Scheithauer$^{3}$, J.~Shangguan${^8}$, J.~Stadler${^8}$, O. Straub${^8}$, C.~Straubmeier$^{4}$, E.~Sturm$^{8}$, E. van Dishoeck$^{8,14}$, F.~Vincent$^{9}$, S. von Fellenberg${^8}$, F.~Widmann$^8$, E.~Wieprecht$^{8}$, M.~Wiest$^{4}$, E.~Wiezorrek$^{8}$, J.~Woillez$^{7}$, S.~Yazici$^{8,4}$ \& G.~Zins$^{10}$
}

\begin{document}

\maketitle

\begin{affiliations}
  \item School of Physics, University College Dublin, Belfield, Dublin 4, Ireland 
  \item Dublin Institute for Advanced Studies, 31 Fitzwilliam Place, D02\,XF86 Dublin, Ireland
  \item Max Planck Institute for Astronomy, K\"{o}nigstuhl 17, Heidelberg, Germany, D-69117
 \item I. Physikalisches Institut, Universität zu Köln, Zülpicher Str. 77, 50937, K\"{o}ln, Germany
 \item Faculdade de Engenharia, Universidade do Porto, Rua Dr. Roberto Frias, P-4200-465 Porto, Portugal
 \item CENTRA, Centro de Astrofísica e Gravitação, Instituto Superior Técnico, Avenida Rovisco Pais 1, P-1049 Lisboa, Portugal
 \item European Southern Observatory, Karl-Schwarzschild-Str. 2, 85748 Garching, Germany
 \item Max Planck Institute for Extraterrestrial Physics, Giessenbachstrasse, 85741 Garching bei M\"{u}nchen, Germany
 \item LESIA, Observatoire de Paris, Université PSL, CNRS, Sorbonne Université, Université de Paris, 5 place Jules Janssen, 92195 Meudon, France
 \item European Southern Observatory, Casilla 19001, Santiago 19, Chile
 \item Max-Planck-Institute for Radio Astronomy, Auf dem H\"{u}gel 69, 53121 Bonn, Germany
 \item Department of Physics, Le Conte Hall, University of California, Berkeley, CA 94720, USA
 \item Unidad Mixta Internacional Franco-Chilena de Astronomía (CNRS UMI 3386), Departamento de Astronomía, Universidad de Chile, Camino El Observatorio 33, Las Condes, Santiago, Chile
 \item Sterrewacht Leiden, Leiden University, Postbus 9513, 2300 RA Leiden, The Netherlands
 \item Univ. Grenoble Alpes, CNRS, IPAG, F-38000 Grenoble, France
 \item Department of Space, Earth \& Environment, Chalmers University of Technology, SE-412 93 Gothenburg, Sweden
 \item Universidade de Lisboa - Faculdade de Ciências, Campo Grande, P-1749-016 Lisboa, Portugal
 
 \item Instituto de Astronom\'{i}a, Universidad Nacional Aut\'{o}noma de M\'{e}xico, Apdo. Postal 70264, Ciudad de M\'{e}xico, 04510, M\'{e}xico 
 
\end{affiliations}

\begin{abstract}

Stars form by accreting material from their surrounding disks. There is a consensus that matter flowing through the disk is channelled onto the stellar surface by the stellar magnetic field. This is thought to be strong enough to truncate the disk close to the so-called corotation radius where the disk rotates at the same rate as the star.
Spectro-interferometric studies in young stellar objects show that Hydrogen is mostly emitted in a region of a few milliarcseconds across, usually located within the dust sublimation radius\cite{kraus08, eisner_tt07, perraut16}. Its origin is still a matter of debate and it can be interpreted as coming from the stellar magnetosphere, a rotating wind or a disk. In the case of intermediate-mass Herbig AeBe stars, the fact that the Br$\gamma$\ emission is spatially resolved rules out that most of the emission comes from the magnetosphere\cite{kurosawa16, rebeca15, ale15}. This is due to the weak magnetic fields (some tenths of G) detected in these sources\cite{hubrig15, alecian13}, resulting in very compact magnetospheres. In the case of T Tauri sources, their larger magnetospheres should make them easier to resolve. However, the small angular size of the magnetosphere (a few tenths of milliarcseconds), along with the presence of winds\cite{banzatti19,simon16} emitting in Hydrogen make the observations interpretation challenging.
Here, we present direct evidence of magnetospheric accretion by spatially resolving the inner disk of the 60\,pc\cite{gaia} T Tauri star TW Hydrae through optical long baseline interferometry. We find that the hydrogen near-infrared emission comes from a region approximately 3.5 stellar radii (R$_{*}$) across. This region is within the continuum dusty disk emitting region (R$_{cont}$ = 7 R$_{*}$) and smaller than the corotation radius which is twice as big.
This indicates that the hydrogen emission originates at the accretion columns, as expected in magnetospheric accretion models, rather than in a wind emitted at much larger distance ($>$1\,au). 
\end{abstract}

TW Hya belongs to an association of young stars around 8\,Myr old. Its proximity to Earth, as well as its favourable pole-on orientation\cite{huang18}, makes it an ideal candidate for protoplanetary disk studies. The disk structure of TW Hya includes a dust-depleted inner hole, as well as a series of bright rings, the closest one located at $\sim$1\,au from the star\cite{huang18, van_Boekel17}. The presence of the inner hole as well as the small near-IR excess\cite{calvet00, manara14} made TW Hya the prototypical “transitional disk” where planets and/or photoevaporation are expected to be the main mechanism of disk dispersal. However, the measurement of non-negligible  accretion rates (2.3$\times$10$^{-9}$\,M$_{\odot}$\,yr$^{-1}$)\cite{venuti19} indicates that the inner disk region of TW Hya is still rich in gas. Further evidence of accretion is given by the detection of a near pole-on cool photospheric spot (stable over several years), coincident with the location of the main magnetic pole (B $\approx$ 2.5\,kG), and a region of accretion-powered excess line emission\cite{donati11}. This suggests that accretion in TW Hya takes place mostly poleward, and that the stellar magnetic field is strong enough to magnetically truncate the inner disk at a few stellar radii from the star. This value is equivalent to a few tenths of milli-arcsecond (mas), and thus it is impossible to directly resolve the magnetospheric accretion region, even for such a nearby star, using conventional methods. This leaves spectro-interferometry as the only suitable technique up to the task. 

With this aim, we conducted high-angular resolution observations of the hydrogen Br$\gamma$\ line in TW Hya using the Very Large Telescope Interferometer (VLTI) instrument GRAVITY with the four 8-m Unit Telescopes (UTs; Fig.\,1). The Br$\gamma$\ line is a well-known tracer of accretion in low-mass protostars through an empirical relationship that relates the line and accretion luminosities\cite{muzerolle98, alcala14}. Our interferometric measurements allowed us to probe the Br$\gamma$\ line and K-band emitting regions along six different baselines (projected baselines ranging from $\sim$130\,m to $\sim$45\,m, resulting in nominal angular resolutions of $\sim$4\,mas to $\sim$10\,mas) and at various position angles. By fitting a geometrical model (see Methods) to the continuum emission (star plus continuum circumstellar emission) and assuming a K-band to stellar flux ratio of 1.18\cite{manara14,venuti19}, we derive a stellar radius of R$_*$ = 1.29 $\pm$ 0.19\,R$_{\odot}$\ (consistent with theoretical expectations) and a radius for the K-band continuum excess/circumstellar emission of  R$_{circ}$ = 6.50 $\pm$ 0.16\,R$_*$ (see Fig.\,2). These values are in agreement with previous interferometric results, and spectroscopic studies\cite{akeson11,manara14, venuti19}. Furthermore, the location of the K-band excess emission is consistent with the location of a disk rim due to silicate sublimation (see Methods).

By removing the continuum contribution to the line emission (see Methods), we find that the Br$\gamma$\ line emitting region is very compact, but nonetheless, marginally resolved for the longest projected baselines (PBL$\gtrsim$60\,m). This allows us to measure a radius for the Br$\gamma$\ emitting region of R$_{Br\gamma}$ = 3.49 $\pm$ 0.20\,R$_*$  assuming a distance of $\sim$60\,pc to TW Hya (see Fig.\,2). This size is consistent with the small $\lesssim$1$^o$ ($\lesssim$2$^o$ total amplitude) photocentre shift of the line with respect to the continuum (the so-called differential phase) detected in our longest baselines (see Fig.\,1). Such a differential phase roughly translates into a Br$\gamma$\ line displacement of $\lesssim$5\,R$_*$ (see Methods for more details), in agreement with the value derived from the continuum-subtracted Br$\gamma$\ line visibilities. 

The inferred size of the Br$\gamma$\ line emission is too compact to be emitted in a photoevaporative wind that in TW Hya is expected to be launched beyond the dust cavity (R $>$ 0.5-1\,au, i.e., R $>$ 80-160\,R$_{*}$)\cite{pascucci11,ercolano17}. 
It should be pointed out that in TW Hya there is no evidence of the presence of a disk wind, which is typically emitted within 0.5\,au from the source, or a jet, which would be associated with bright fast blue-shifted emission in lines such as H$\alpha$, and [OI]\,6300\,\AA, [SII]\,6717\,\AA, that is not observed in this object\cite{fang18,banzatti19}. 
Therefore, the results presented here indicate that the Br$\gamma$\ line is emitted in the magnetospheric accretion region. Classical magnetospheric accretion models assuming free fall velocities along an axisymmetric, dipolar magnetosphere predict indeed that the Br$\gamma$\ line is formed along the accretion columns\cite{muzerolle01,kurosawa11, gullbring98}. In these models, the Br$\gamma$\ line has a broad profile, comparable to the free-fall velocity, and centred around zero velocity. This is the case of the Br$\gamma$\ line observed in TW Hya that shows a full width at zero intensity (FWZI) of $\sim$ 400\,km/s consistent with the expected velocity of gas around a solar mass star falling at free-fall from $\sim$3-4\,R$_{*}$. Therefore, our measurements indicate that the Br$\gamma$\ line is emitted along the magnetospheric accretion columns that truncate the disk at around 3.5\,R$_{*}$. 

Is this value consistent with the expected magnetospheric truncation radius of TW Hya as determined by its magnetic field? Zeeman-Doppler imaging has been used to reconstruct the magnetic field topology and strength in TW Hya\cite{donati11}. These measurements showed that the magnetic field of TW Hya is strong ($\sim$1.5\,kG) and mostly poloidal and axisymmetric with respect to the stellar rotation axis. The field can be separated into a complex $\sim$2.5\,kG octupole component and a much fainter $\sim$400-700\,G dipolar large-scale field. Models for such complex magnetic field topologies show that the gas initially accretes following the dipolar field lines, although near the stellar surface the octupole component alters the flow of matter \cite{gregory08,romanova11}. Following this idea, and the theoretical work of Bessolaz et al.\cite{bessolaz08}, we estimate a truncation radius of 3-4\,R$_{*}$\ assuming a stellar radius and mass of 1.22\,R$_{\odot}$\ and 0.6\,M$_{\odot}$\cite{venuti19}, and a mass accretion rate of 2.3$\times$10$^{-9}$\,M$_\odot$/yr\cite{manara14,venuti19} for TW Hya, and a strength of the dipolar magnetic field component of 400-700\,G. Therefore, the size of the Br$\gamma$\ line emitting region derived from our interferometric measurements and the size of the truncation radius estimated from the magnetic field of TW Hya are strikingly similar.
In addition, the measured size of the line emission is inconsistent with a disk wind, since it is significantly smaller than the inferred truncation radius. There is a small possibility that dust-free disk gas extends inwards of the inferred sublimation radius, and could be responsible for at least part of the line emission. However, the previously measured magnetic field strength and geometry implies a disk truncation radius consistent with the K-band continuum size. Finally,  the detection of spatially-resolved line emission rules out that most of the Br$\gamma$\ emission is originated at the accretion shock near the stellar surface. A schematic view of our findings is shown in Fig.\,2 in Methods.
Our results are then in agreement with the topology and strength of the magnetic field and they validate the assumption that when the magnetic field of the central star is complex, the truncation radius is located closer to the central star than would be expected if the magnetic field has a dipolar morphology of similar average strength\cite{gregory08, johnstone14}.






\section*{Bibliography}


%
%
\begin{table}
\caption{{\bf Size estimates derived from the best fit of the continuum and continuum-compensated Br$\gamma$\ line data.} }
\label{tab:fits}
\centering
\begin{tabular}{c c c c}
\hline 
TW\,Hya & R & R & R  \\ 
      & [mas]  & [au]  & [R$_{\odot}$]  \\
      \hline
star  & 0.10 $\pm$ 0.01 & 0.006 $\pm$ 0.001 & 1.29 $\pm$ 0.19 \\ 
disk  & 0.65 $\pm$ 0.02 & 0.039 $\pm$ 0.001 & 8.39 $\pm$ 0.21  \\
line  & 0.35 $\pm$ 0.02 & 0.021 $\pm$ 0.001 & 4.50 $\pm$ 0.26  \\
\hline
\end{tabular}
\end{table}
\
\newline
\begin{figure}
\centering
	\includegraphics[width=183mm]{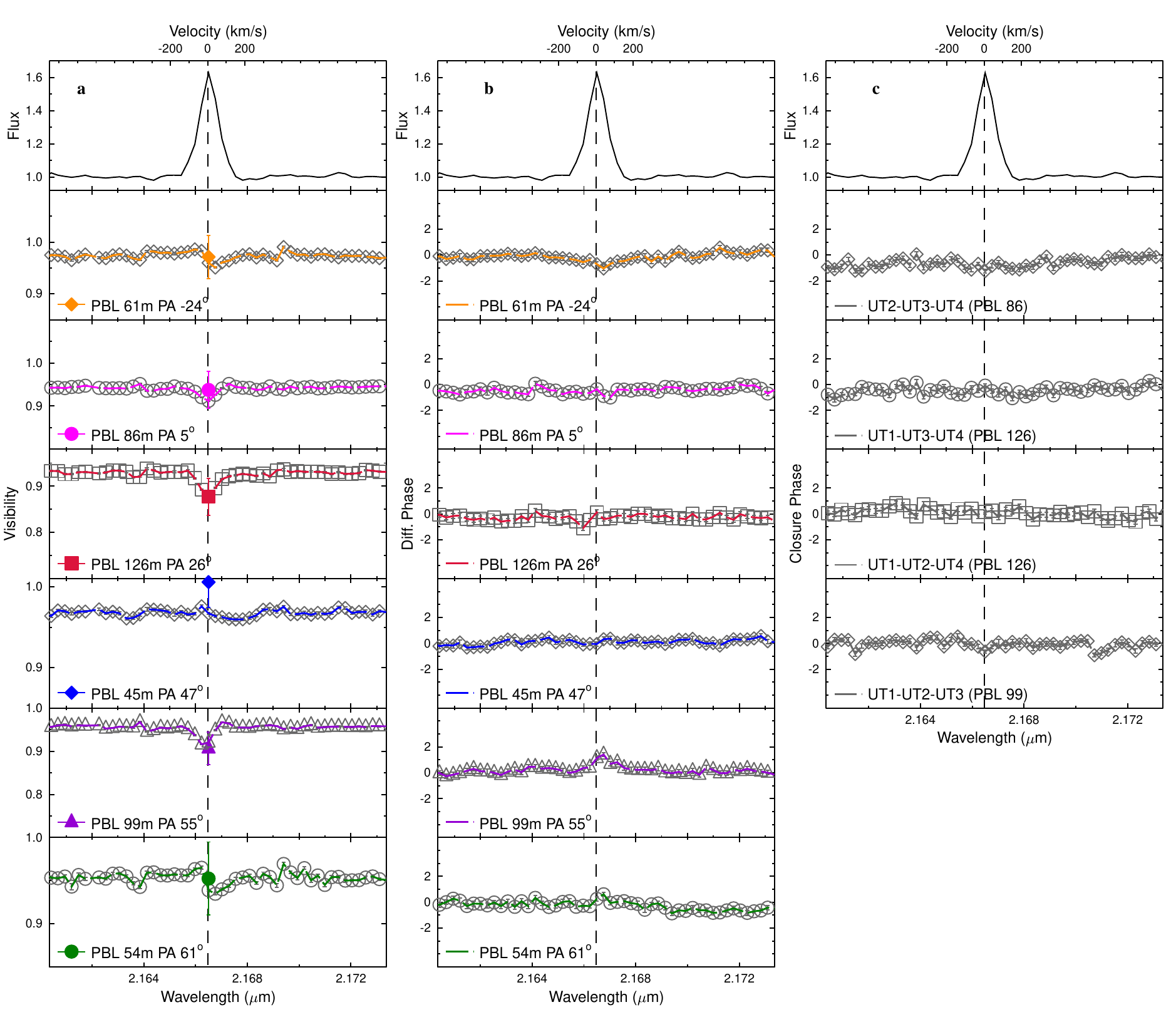}
	\caption{{\bf VLTI-GRAVITY observations of TW\,Hya.} The Br$\gamma$\ line profile normalised to the continuum is shown in the top row. The radial velocity is with respect to the stellar reference. {\bf a:} Wavelength-dispersed visibility amplitudes for the six baselines along with the errors derived from the data reduction. The continuum-subtracted Br$\gamma$\ line visibilities are shown as full coloured symbols. The associated errors were derived by propagating Eq. 1 in Methods as described in the text. {\bf b:} Same as ${\rm a}$, but for the wavelength-dispersed differential-phase signals. {\bf c:}  Wavelength-dispersed closure phase signals for the triplets UT2-UT3-UT4, UT1-UT3-UT4, UT1-UT2-UT4 and UT1-UT2-UT3. The maximum PBL of each baseline is indicated in between parenthesis.}
\end{figure}
%
%
\newline
%
%
\begin{figure}
	\includegraphics[width=89mm]{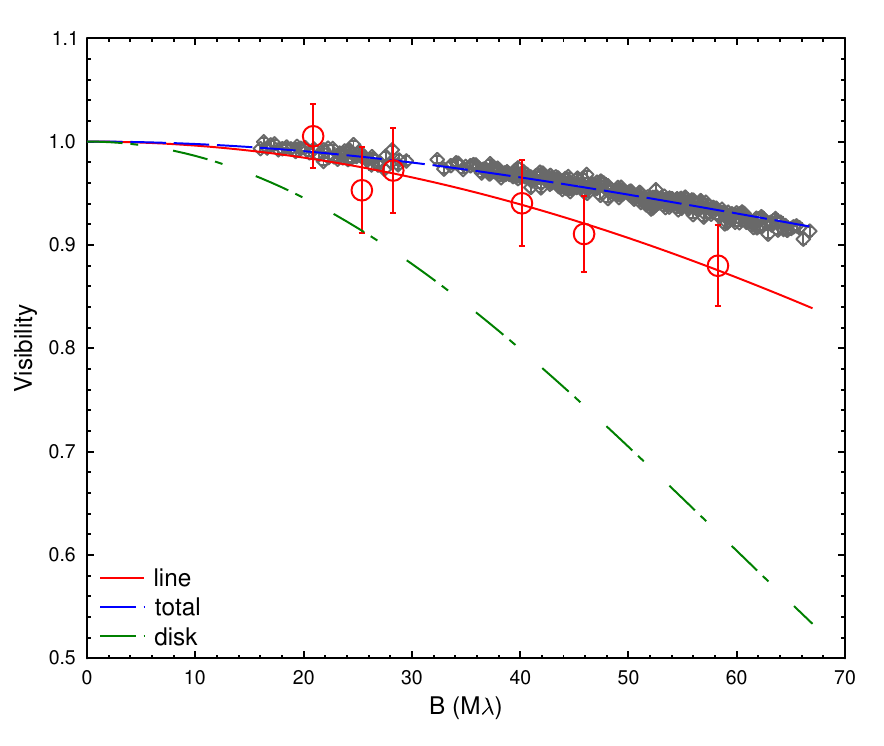}
	\caption{{\bf Visibility plot of TW\,Hya.} Visibility points versus megalambda for the continuum calibrated FT data (open grey diamonds) and the continuum-subtracted Br$\gamma$\ line (open red circles). The fit to the total continuum and continuum-subtracted Br$\gamma$\ visibilities is shown in dashed blue and solid red lines. The circumstellar/disk visibility curve is also shown in dashed-dotted green line as reference.}
    \label{fig:fit}
\end{figure}
%



\begin{methods}

\section{Observations and data reduction}


TW\,Hya\ was observed with the VLTI instrument GRAVITY\cite{GRAVITY} on 21 January 2019 using the four 8m Unit Telescopes (UTs) of the European Southern Observatory (see Table\,1 in Extended Data). The target was observed in single-field combined polarisation mode (i.e. fringes were tracked and servoed on the target itself) and operating with the MACAO on-axis adaptive optics system. The data on the fringe tracker (FT) detector were recorded at low spectral resolution (${\rm R}$~$\sim$~23) with a detector integration time per interferogram (DIT) of 0.85\,ms, whereas the science (SC) detector was working at high spectral resolution (HR; ${\rm R}$~$\sim$~4000, i.e., $\Delta {V} \sim$70\,km\,s$^{-1}$) and with a DIT of 30\,s. 

The data were reduced using the GRAVITY pipeline version 1.3.0\cite{DRS}. The atmospheric transfer function was calibrated using the calibrators HD\,91937 and HD\,95470 (see Table\,1 in Extended Data). The spectrum of TW\,Hya\ was obtained by averaging the HR spectra recorded in the four photometric channels. Standard telluric correction was also applied to the spectrum using HD\,95470 (SpT K2/3\,III) as a telluric standard star. Finally, the spectrum was flux calibrated assuming a 2MASS K-band magnitude of 7.3 for TW Hya. The wavelength calibration of the spectra was refined using several telluric absorption lines present in the spectrum. An average shift of $\sim$4\AA\ was applied to the data.  

\section{Interferometric Observables}

VLTI-GRAVITY observations of TW\,Hya\ provided us with the K-band spectrum of the source, six spectrally dispersed visibilities (that give a measure of the size of the object, with V=1 indicating a point source, and V=0 indicating a fully resolved object) and differential phases (that measure the photocenter shift of the line with respect to the continuum) and four closure phases (that provides a measure of the asymmetry of the continuum and/or line emission) (see Fig.\,1 in the print paper). 

The spectrum of TW\,Hya\ shows bright Br$\gamma$\,2.166\,$\mu$m\ line emission, along with NaI\,2.206\,$\mu$m, and NaI\,2.209\,$\mu$m, and rovibrational CO in absorption. No interferometric signal is detected for any of these lines except the Br$\gamma$\ line. A small differential phase signature of 2$^o$ in the Br$\gamma$\ line is detected along the two longest baselines. No closure phases were detected within the errors. 

The continuum visibilities point towards a very compact circumstellar environment around TW\,Hya\ with measured continuum visibilities above $\sim$0.95 in all our baselines. Interestingly, the total visibilities within the line decreases with respect to the continuum visibilities, indicating that the sum of the Br$\gamma$\ emitting region plus the continuum contribution (including the stellar plus circumstellar environment) is more extended than the continuum alone. However, it should be noted that the total visibility is not just the sum of each visibility component but it is weighted by the flux of each component. In other words, assuming that the level of the continuum within and outside the line is the same and that the differential phase is zero then: V$_{tot}$ F$_{tot}$ = V$_{cont}$ F$_{cont}$ + V$_{line}$ F$_{line}$, with V$_{cont}$ F$_{cont}$=V$_*$ F$_*$ + V$_{circ}$ F$_{circ}$; and  F$_{tot}$ = F$_{cont}$ + F$_{line}$. In these expressions, F$_{tot}$, F$_{cont}$, F$_{line}$, and V$_{tot}$, V$_{cont}$, V$_{line}$ are the total, continuum and line fluxes, and visibilities respectively; and F$_{*}$, F$_{circ}$, and V$_*$, V$_{circ}$ are the stellar and circumstellar continuum fluxes, and visibilities.
In order to further investigate the circumstellar and Br$\gamma$\ line emitting region the contribution from the star and the overall continuum emission must be removed from the measured visibilities.

\section{Circumstellar continuum emitting region}

As mentioned above, in order to estimate the size of the continuum circumstellar emitting region the emission from the star must be removed. In doing so, a K-band to stellar flux ratio of 1.18 was assumed\cite{manara14}. As we expect to marginally resolve TW\,Hya\ due to its close distance (i.e. V$_{*}<$1 on our longest baselines), we took the conservative approach of fitting two Gaussian components to the continuum visibilities measured by GRAVITY, one corresponding to the central star plus an additional component due to the circumstellar disk (assumed to be inclined, that is, the inclination and position angle were allowed to be free paramenters). 
Using this approach, our best fitting model corresponds to two Gaussians with full-width-at-half-maximum (FWHM) of FWHM$_*$ = 0.20 $\pm$ 0.03\,mas and FWHM$_{circ}$ = 1.30 $\pm$ 0.04\,mas, respectively. The inclination ($i$) and position angle (PA) of the latter component is consistent with a nearly face-on structure as reported by ALMA\cite{huang18}. However, due to the lack of long baselines, along with the nearly face-on geometry, we cannot give stringent constraints on the $i$ and PA of the system, and from now on we will assume ALMA measurements of $i\sim$5$^o$\ and PA$\sim$32$^o$\ as our fiducial values.
Coming back to the size of the emitting region, the derived FWHM values correspond to a stellar radius of R$_*$ = 1.29 $\pm$ 0.19\,R$_{\odot}$\ and R$_{circ}$ = 8.39 $\pm$ 0.21\,R$_{\odot}$\ assuming a distance of 60\,pc. The retrieved stellar and circumstellar radii are in good agreement with previous values found in the literature\cite{eisner06, akeson11, manara14,sokal18}.   
If a lower value of the observed K-band to stellar flux ratio of 1.10 is assumed\cite{eisner10_TWHya, vacca11}, it would provide a worse fit, with stellar and circumstellar radii with much larger errors, namely, R$_*$= 0.05$\pm$0.11\,mas (i.e., 0.68$\pm$1.42\,R$_\odot$), and R$_{circ}$= 0.70$\pm$0.13\,mas (i.e. 9.03$\pm$1.68\,R$_{\odot}$).

\section{Continuum-subtracted Br$\gamma$\ line visibilities}
\label{methods:size_BrG}

The size of the Br$\gamma$\ line emitting region can be estimated by assuming that the total visibilities within the Br$\gamma$\ line are due to the contribution of the line emitting region plus the continuum component. In this way, the pure (or continuum compensated) Br$\gamma$\ line visibilities can be derived by subtracting the continuum contribution from the total line visibilities following\cite{weigelt07}:

\begin{equation}
V_{cont} V_{tot} e^{i\Phi'} = \frac{V_{cont}}{F_{cont} + F_{line}} (F_{cont} V_{cont} + F_{line} V_{line} e^{i\Delta \Phi})
\end{equation}
where $\phi'$ is the differential phase in the line, and $\Delta\Phi$ is the difference of the Fourier phases of  the  continuum  and  line  components, that is $\Delta\Phi (B/\lambda_{line}) = \Phi_{cont}(B/\lambda_{line}) - \Phi_{line}(B/\lambda_{line})$. 
Thus the errors on the continuum compensated visibilities have been estimated taking into account the error on the continuum and total visibilities (assuming the rms value as a conservative error), and the differential phase, $\phi$, errors.

Initially, the continuum-compensated Br$\gamma$\ line visibilities were computed at three velocity channels at radial velocities of $\sim$-33\,km\,s$^{-1}$, $\sim$4\,km\,s$^{-1}$, and $\sim$40\,km\,s$^{-1}$, and with a line-to-continuum ratio higher than 10\%. 
For all six baselines, the continuum-compensated Br$\gamma$\ line visibilities measured at each spectral channel are roughly the same within the errors. Therefore, the weighted-mean of the three pure line visibilities for each baseline was computed and used to derive the size of the Br$\gamma$\ line emitting region. The average pure Br$\gamma$\ line visibilities are shown in Fig.\,1 and 2 in the print paper. The Br$\gamma$\ line emitting region is marginally resolved only for the longest PBLs (PBL $\gtrsim$ 60\,m), meaning that the emitting region is very compact.

To derive the size of the Br$\gamma$\ line emitting region, we computed a geometric model of the Br$\gamma$\ line continuum-compensated visibilities using a Gaussian fit. As for the continuum, we fixed the inclination and position angle to the values derived by ALMA and we fitted the line visibilities with only the Gaussian FWHM as a free parameter. The best fit result is shown in Table\,1 in the print paper, and it corresponds to a radius of the Br$\gamma$\ line emitting region of R$_{Br\gamma}$ = 0.35 $\pm$ 0.02\,mas or R$_{Br\gamma}$ = 4.5 $\pm$ 0.26\,R$_{\odot}$, assuming a distance of $\sim$60\,pc to TW\,Hya.

In order to probe the effect of the assumed F$_{tot}$/F${_{cont}}$ flux ratio on our results, we have repeated the analysis varying this ratio by 10\%. The results (R$_{Br\gamma}^{-10\%}$ = 0.37 $\pm$ 0.05\,mas; R$_{Br\gamma}^{-10\%}$ = 0.33 $\pm$ 0.02\,mas) are consistent with the previous one within the error bars. 

\section{Continuum-subtracted Br$\gamma$\ line differential phase}

As for the case of the visibilities, the contribution of the continuum to the differential phase can be removed. This type of analysis is specially useful when the measured photocentre shift of the line is weak\cite{whelan09, vane14}. Following \cite{weigelt07}, the displacement of the photocentre of the line at any given wavelength can be derived from:
\begin{equation}
sin(\Delta\phi)=sin(\phi)\cdot \dfrac{|F_{tot}V_{tot}|}{|F_{line}V_{line}|}
\end{equation}
The displacement of the photocenter of the emission at any given wavelength $\delta$ is then:
\begin{equation}
\delta = -\Delta \phi {{\lambda}\over{2 \pi B}}
\end{equation}
where $B$ is the length of the baseline. The upper limit of the differential phase is $\sim$1\degree. This translates into a maximum 
value of $\Delta \phi_{max}^{Br\gamma}$=-6.2\degree, equivalent to a maximum displacement of $\delta_{max}\sim$3.8\,R$_{\odot}$ $\sim$ 4.9\,R$_*$.
This value is very similar to the one derived from the continuum-subtracted visibilities.

\section{Rim Radius}

We can estimate the rim radius (or the distance from the star where silicates sublimates) using eq.11 of Dullemond et al.\cite{dullemond01} under the assumption that the pressure scale height is a small fraction of R.  For the temperature (T$\sim$4000\,K) and radius (R$\sim$1.2R$_{\odot}$) assumed here for TW Hya, a rim radius of $R_{rim}\sim 7.5 R_{*}$ is found for a sublimation temperature T$_{sub}$=1500\,K.

\section{Corotation radius}

The corotation radius depends on the stellar mass, radius, and rotation velocity. This latter is uncertain due to the low inclination of TW\,Hya with respect to the line of sight. Estimates of the rotation velocity ranges from 80 $\pm$ 34\,km/s (assuming a $v \sin i =$ 7 $\pm$ 3\,km/s\cite{venuti19} and a disk inclination of 5$^o$ \cite{huang18}) to $\sim$ 17.4\,km/s (assuming a rotation period of $P=3.56$\,d\cite{setiawan08}). Taking these values, and a stellar mass and radius of 0.58\,M$_{\odot}$\ and 1.22\,R$_{\odot}$\ \cite{venuti19}, we find a corotation radius of  R$_{co}$ $\sim$ 6.5 -- 7.1\,R$_{*}$. The corotation radius is significantly larger than R$_{Br\gamma}$, supporting our hypothesis that the Br$\gamma$\ size measures the radius of the magnetosphere, most likely tracing the width of the region containing accretion columns.

\end{methods}

\section*{Bibliography}
\vspace{1cm}

\newcounter{mybibstartvalue}
\setcounter{mybibstartvalue}{30}

\xpatchcmd{\thebibliography}{%
  \usecounter{enumiv}%
}{%
  \usecounter{enumiv}%
  \setcounter{enumiv}{\value{mybibstartvalue}}%
}{}{}




\begin{addendum}
    
    \item[Data availability] Based on observations collected at the European Southern Observatory under ESO programme 0102.C-0408(C). The raw data are publicly available in the ESO Science Archive Facility. The reduced data that support the findings of this study are available from the corresponding author under reasonable request. 

 \item  The authors would like to thank Dr. C. Manara for kindly providing the XSHOOTER spectrum of TW Hya and the template of the stellar photosphere. This  material  is  based  upon  works  supported  by  Science  Foundation  Ireland under Grant No. 18/SIRG/5597. M.K. was funded by the Irish Research Council (IRC), grant GOIPG/2016/769. R.F. acknowledges support from Chalmers Initiative on Cosmic Origins (CICO) postdoctoral fellowship. A.C.G. and T.P.R. have received funding from the European Research Council (ERC) under the European Union's Horizon 2020 research and innovation programme (grant agreement No.\ 743029). A.N. acknowledges the kind hospitality of DIAS. A.A., M.F. and P.J.V.G acknowledge funding by Fundação para a Ciência e a Tecnologia, with grants reference UID/FIS/00099/2013 and SFRH/BSAB/142940/2018. T.H. acknowledges support from the European Research Council under the Horizon 2020 Framework Program via the ERC Advanced Grant Origins 832428. 

\item[Authors contributions] GRAVITY is developed in a collaboration by the Max Planck Institute for Extraterrestrial Physics, LESIA of Paris Observatory and IPAG of Université Grenoble Alpes / CNRS, the Max Planck Institute for Astronomy, the University of Cologne, the Centro Multidisciplinar de Astrofisica Lisbon and Porto, and the European Southern Observatory. Authors from these institutes 
contributed to the concept, design, assembly, instrumental tests, science cases, verification and implementation of GRAVITY and its subsystems, and the data reduction pipeline. 
P.J.V.G. conducted the observations. R.G.L., K.P. and M.K. reduced the data. R.G.L. and M.K. analysed the data. A.N. estimated the location of the disk rim, and the corotation and truncation radius. R.F. performed the model fitting of the continuum-subtracted visibilities. R.G.L. wrote the manuscript. A.N., T.P., A.C.G. edited the manuscript.  R.G.L., A.N., A.C.G., T.P.R., R.F., M.K., L.K., K.P., J.S.-B., M.B., C.D., L.L., W.B., P.J.V.G., T.H., P.C., G.D., T.Z., and R.G discussed the results and their implications, and commented on the manuscript. 

 \item[Competing Interests] The authors declare that they have no
competing financial interests.
 \item[Correspondence] Correspondence and requests for materials
should be addressed to R.G.L.~(email: rebeca.garcialopez@ucd.ie).
\end{addendum}


\end{document}